\documentstyle[pra,aps,amsfonts,graphicx]{revtex}

\begin{document}
\draft \twocolumn[\hsize\textwidth\columnwidth\hsize\csname
@twocolumnfalse\endcsname
\title{Experimental investigation of quantum key distribution with position and momentum of photon pairs}
\author{M. P. Almeida, S. P. Walborn  and P. H. Souto Ribeiro$^{*}$}
\address{Instituto de F\'{\i}sica, Universidade Federal do Rio de
Janeiro, Caixa Postal 68528, Rio de Janeiro, RJ 21941-972, Brazil}
\date{\today}
\maketitle
\begin{abstract}
We investigate the utility of Einstein-Podolsky-Rosen correlations of the
position and momentum of photon pairs from 
parametric down-conversion in the implementation of a secure quantum key
distribution protocol. We show that security is guaranteed by the entanglement
between down-converted pairs, and can be checked by either direct comparison of Alice and Bob's measurement results or 
evaluation of an inequality of the sort proposed by Mancini {\it et al.} (Phys. Rev. Lett. {\bf 88}, 120401 (2002)).   
\end{abstract}
\pacs{42.50.Ar, 42.25.Kb} ]
\section{Introduction}
Quantum communication protocols using photonic qubits have been proposed
and implemented, utilizing entanglement in several degrees of freedom of the
photon\cite{bouwmeester00}.  Perhaps the most promising application of quantum communication using photons is quantum key distribution (QKD) \cite{bouwmeester00,1}, in which entangled or single qubits are sent from sender to receiver(s) and used to establish a secret random key string, which can then be used to securely transmit a data string.  Most QKD schemes are inspired by the original single-qubit BB84 \cite{bb84} or entangled-qubit Ekert \cite{ekert} protocols.  In the BB84 protocol, cryptographic security is provided by the partial indistinguishability of non-orthogonal states and the no-cloning theorem, while the security of the Ekert protocol is guaranteed by violation of Bell's inequality.    
For photon pairs, obtained from spontaneous parametric down-conversion (SPDC), entanglement in polarization \cite{3} and  energy-time \cite{4} have been most widely and sucessfully used.
\par  
There have also been QKD proposals based on continuous variable field quadratures of multiphoton beams \cite{ralph99}.  Again, security of these protocols is based on either measurement of non-commuting observables and the no-cloning theorem or 
violation of some classical inequality.
\par 
  
\par 
In this paper, we present a protocol for QKD based on the position and momentum degrees of freedom entangled photons created by SPDC.  Recently, it has been shown that the difference between the positions
of entangled SPDC photons along with the sum of their momenta are Einstein-Podolsky-Rosen (EPR) correlated,
or, in other words, entangled\cite{5}.  These correlations are interesting not only for their direct relation to the seminal paper by EPR \cite{epr35}, but also for their possible application to quantum information tasks. To our knowledge, the work we present here is the first quantum information protocol based on position and momentum entanglement of the form originally proposed by EPR.  Entanglement in position and momentum is
easily obtained, since it is a direct consequence of the inherent phase matching
conditions in the parametric process. Moreover, it has been demonstrated that it is possible to protect this type of entanglement against divergence due to free-space transmission of the down-converted photons \cite{almeida}.  Therefore quantum key distribution in position-momentum variables could be  a promising
application. 
\par
In section \ref{sec:theory} we present the QKD protocol, and discuss basic security issues in section \ref{sec:security}.  Section \ref{sec:experiment} shows experimental results confirming the utility of position and momentum EPR correlations in QKD.  
\section{Theory}
\label{sec:theory}
\begin{figure}[h]
\includegraphics*[width=8cm]{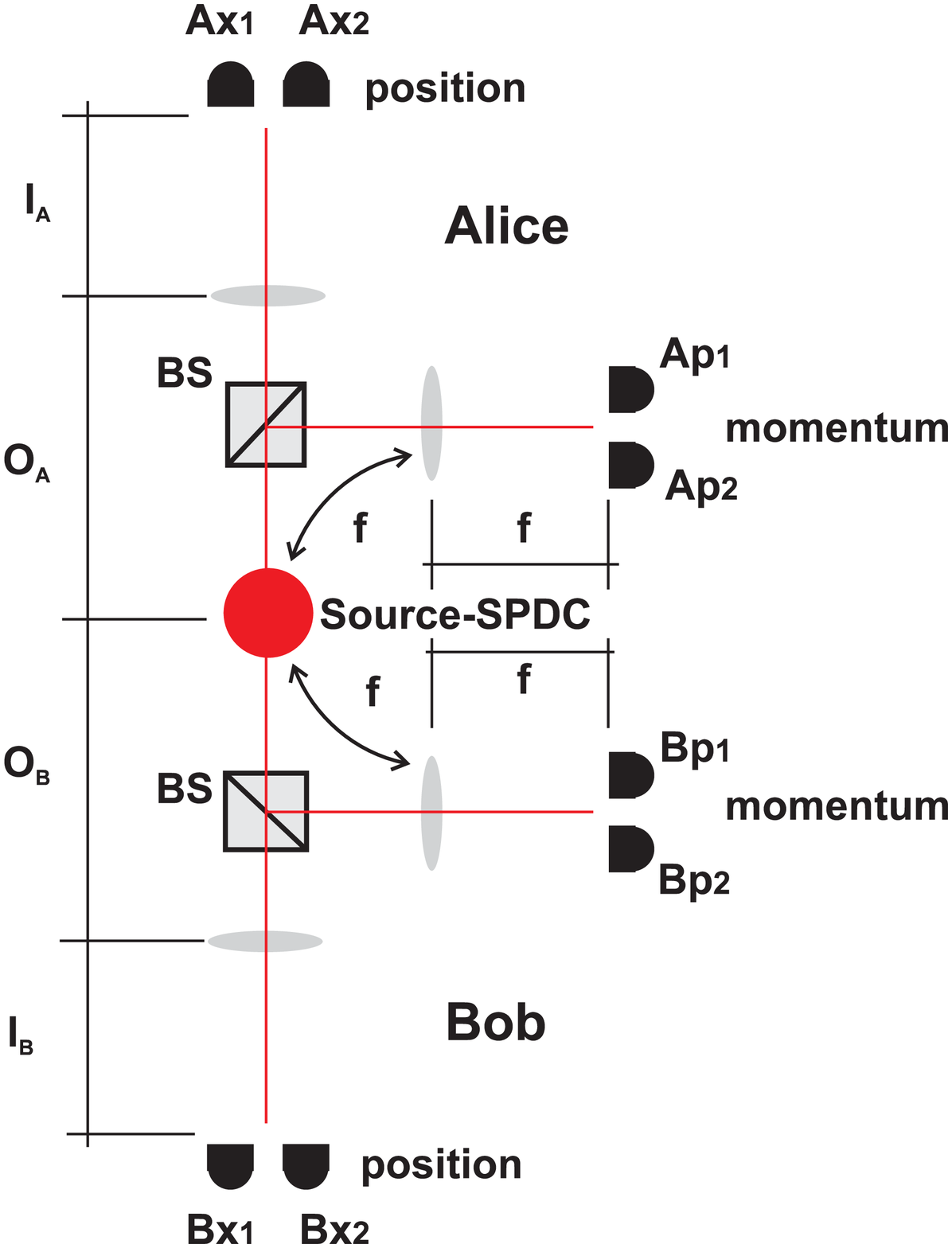}
\vspace*{0.25cm} \caption{Diagram of quantum key distribution using position and momentum variables of down-converted photons.  The SPDC source emits photons entangled in position and momentum.  Lenses are used to create the image (position measurments) or Fourier transform (momentum measurements) in Alice and Bob´s detection planes.  THe 50-50 non-polarizing beam splitters (BS) are used to choose randomly between position and momentum measurements.} \label{fig1}
\end{figure}
\par
 Fig. \ref{fig1} shows a diagram of the
basic QKD apparatus.  The protocol works as follows.  Twin photons are created by a SPDC source.  One photon of each pair is sent to Alice (A), and one to Bob (B).  At each of Alice and Bob's stations, a random selection between
detection in the position ($x$) or momentum ($p$) basis is performed by a 50-50 non-polarizing beam splitter (BS).  A measurement in the $x$ basis is
easily performed by  a lens imaging the crystal onto the detection
plane.  It is well known that a measurement in the $p$ basis can be
implemented by a lens with crystal and detection planes in its focal
planes \cite{stoler}, as was done in Refs. \cite{5}.  

\par
Before any photons transmitting potential key bits are exchanged, 
Alice and Bob calibrate their $x$ detectors
for positions A$x_1$ and A$x_2$ and B$x_1$ and B$x_2$, and their $p$ detectors at positions A$p_1$
and A$p_2$ and B$p_1$ and B$p_2$, respectively.  Let us assume that, in both bases, detector position ``1" (``2") represents the 0 (1) logical level.  From the multimode theory of SPDC \cite{almeida,9}, it is possible to show that, for ideal point-like detectors, the probability to detect photon pairs is  
 \begin{equation}
 P_{xx}(\bbox{\rho}_{A},\bbox{\rho}_{B}) \propto \left|{\mathcal{W}}\left(\alpha \bbox{\rho}_{A} + \beta \bbox{\rho}_{B}\right)\delta\left(\alpha \bbox{\rho}_{A} - \beta \bbox{\rho}_{B}\right)\right|^2
 \label{eq:1}
\end{equation}
 for a position-position ($xx$) configuration.  The notation ``$ij$" refers to the situation where Alice measures in the $i$ basis and Bob in the $j$ basis, where $i,j=x,p$.  Here $\mathcal{W}$ is the field profile of the pump beam at the crystal face. We define the parameters $\alpha\equiv O_A/(2I_A)$ and $\beta\equiv O_B/(2I_B)$, which are related to the magnification factors of Alice's and Bob's imaging systems.  We have also assumed that both imaging lenses have been placed so as to obey the thin lens equation. The delta function appears due to the fact that we have considered ideal point-like detectors.  Since the pump beam profile $\mathcal{W}$ can be made much larger than the delta function, it is possible to find positions $\bbox{\rho}_{A}$ and $\bbox{\rho}_{B}$ such that $\mathcal{W}$ is approximately constant.
\par
Similarly, for a momentum-momentum ($pp$) configuration,    
  \begin{equation}
 P_{pp}(\bbox{\rho}_{A},\bbox{\rho}_{B}) \propto \left|v\left(\frac{k_{A}}{f}\bbox{\rho}_{A} + \frac{k_{B}}{f}\bbox{\rho}_{B}\right)\right|^2,
 \label{eq:2}
\end{equation}
  where $v$ is the angular spectrum of the pump beam at the crystal face.  
  If they measure in the same basis, 
Alice and 
Bob can calibrate their detectors so that they see a large correlation in their coincidence measurements.  For example, if Alice detects a photon at A$x_1$, then the probability for Bob to detect a photon at B$x_1$ will be much greater than the probability to detect a photon at B$x_2$.  In general, Alice and Bob can choose positions such that
\begin{eqnarray} 
P(Ax_{1},Bx_{1}) & \approx P(Ax_{2},Bx_{2}) >> 0 \\
P(Ax_{1},Bx_{2}) & \approx P(Ax_{2},Bx_{1}) \approx 0, 
\end{eqnarray}
where $P(A,B)$ is the coincidence detection probability.  
The same is true for measurements in the $p$ basis:  
\begin{eqnarray} 
P(Ap_{1},Bp_{1}) & \approx P(Ap_{2},Bp_{2}) >> 0 \\
P(Ap_{1},Bp_{2}) & \approx P(Ap_{2},Bp_{1}) \approx 0.  
\end{eqnarray}
Moreover, Alice and Bob can adjust their detection systems such that $P(Ax_{1},Bx_{1})  \approx P(Ax_{2},Bx_{2}) 
\approx P(Ap_{1},Bp_{1}) \approx P(Ap_{2},Bp_{2})$.  In the interest of simplicity, we will remove the subscripts on the probabilities $P_{ij}$ whenever redundant.
\par
The security of the QKD protocol is conditioned on the fact that the correlation is
low when the
measurements are performed in different bases, so that very little information is shared.   For example, if Alice detects a photon at A$x_1$, then the probability for Bob to detect a photon at B$p_1$ should be the same as the probability to detect a photon at B$p_2$.
Using the standard theory of SPDC, the coincidence detection probability for an $xp$ configuration is 
 \begin{equation}
 P_{xp}(\bbox{\rho}_{A},\bbox{\rho}_{B}) \propto |{\mathcal{W}}(\alpha \bbox{\rho}_{A})|^2,
 \label{eq:3}
\end{equation}
while a $px$ configuration gives    
 \begin{equation}
 P_{px}(\bbox{\rho}_{A},\bbox{\rho}_{B}) \propto |{\mathcal{W}}(\beta\bbox{ \rho}_{B})|^2.
 \label{eq:4}
\end{equation}
There is no correlation in the detection probabilities (\ref{eq:3}) and (\ref{eq:4}).  Thus, Alice and Bob can choose their detector positions such that 
$P(Ax_{1},Bp_{1})=P(Ax_{1},Bp_{2})=P(Ax_{2},Bp_{1})=P(Ax_{2},Bp_{2})=P(Ap_{1},Bx_{1})=P(Ap_{1},Bx_{2})=P(Ap_{2},Bx_{1})=P(Ap_{2},Bx_{2})$.
\par
To generate a secret key, Alice and Bob perform a series of measurements on a number of photon pairs until they have accumulated $N$ coincidence events, where $N$ depends on the size of the key and the level of security required.  All events in which only one or neither of them detect a photon are discarded.  After the $N$ photon pairs have been detected, Bob, through classical communication, informs Alice of his measurement basis ($x$ or $p$) for each photon.  On average, Alice and Bob measure in the same basis 50\% of  the time.    They keep these photons and discard the rest.  Alice then chooses $m$ of the remaining photon pairs at random and tells Bob to reveal his measurement result.  She then uses these photon pairs to check for an eavesdropper by comparing her measurement results with those of Bob.   The presence of an eavesdropper is registered by a deterioration of the quantum correlation observed in the $xx$ or $pp$ coincidence events.    If the error rate is below a given threshold, then Alice and Bob can be sure that any eavesdropper has obtained an insignificant amount of information about the secret key.  They can then use classical privacy amplification and information reconciliation protocols to increase security and reduce errors in the key \cite{chuang00}.
\par
\section{Security}
\label{sec:security}
We will now discuss the security of this protocol.  There has been much work in security proofs for a wide variety of attacks on QKD systems \cite{1}.  We will limit our discussion to incoherent attacks, in which the eavesdropper (Eve) has access to one photon pair at a time.  Here we provide a security argument for a simple attack, and leave more complex attacks for future work.  
\par  
If Eve steals one or both photons, then no coincidence is detected, the event is discarded and 
no information is obtained by Eve.  One way for Eve to obtain information is based on an intercept-resend type strategy \cite{1}, in which she steals one or both photons, measures (and thus destroys) them, and then replaces them with new photons.  We note here that we call this type of general attack, in which Eve can measure in any basis, as an intercept-resend ``type" strategy, in contrast to the usual intercept-resend strategy found in the literature in which Eve measures in Alice and Bob's bases.  As in the BB84 or Ekert protocols using qubits, in this attack Eve's presence is marked by a deterioration of the correlation in Alice and Bob's measurements.  
\par
One method of testing security is through the quantum bit error rate, which can be defined as a function of the ``wrong" and ``right" detection probabilities \cite{1}:
 \begin{equation}
 QBER = \frac{P^{\,\mathrm{wrong}}}{P^{\,\mathrm{wrong}}+P^{\,\mathrm{right}}}.
 \label{eq:qber1}
 \end{equation}
In order to simplify the analysis of our experimental results, we will not assume that $P^{\,\mathrm{wrong}}+P^{\,\mathrm{right}}=1$. 
With no eavesdropper present, in our notation the QBER is given by 
 \begin{equation}
 QBER_0 = \frac{\sum\limits_{j=x,p}\sum\limits_{s,t=1,2}^{s\neq t}P({\mathrm{A}}j_s,{\mathrm{B}}j_t)}{\sum\limits_{j=x,p}\sum\limits_{s,t=1,2}P({\mathrm{A}}j_s,{\mathrm{B}}j_t)}.
 \label{eq:qber2}
 \end{equation} 
\par
There are many different attack strategies that Eve can adopt \cite{1}.  
As an example, let us consider that Eve implements the usual intercept-resend strategy on Bob's quantum channel.  Let us suppose that Eve is completely aware of Bob's detection system and has constructed an identical one of her own, and denote $R_{ij}(\rho_{A},\rho_{E})$ as the coincidence detection probability for Alice and Eve, where $i,j=x,p$.    Then, Alice and Eve's detection probability is the same as that of Alice and Bob:  $R_{ij}(\rho_{A},\rho_{E}) = P_{ij}(\rho_{A},\rho_{B})$.  She intercepts and measures Bob's photon, giving one of the following results:  E$x_1$, E$x_2$, E$p_1$, E$p_2$ or a null count.  She then prepares a photon in the eigenstate corresponding her measurement result and sends it to Bob.  We will consider only the cases where Eve detects a photon.  Then, for a given photon pair, Alice, Bob and Eve detect photons with probability
\begin{equation}
P_{ijk}(\rho_A,\rho_B,\rho_E)=R_{ik}(\rho_A,\rho_E)p_{j}(\rho_{B};k),
\end{equation} 
 where  $i,j,k=x$ or $p$, $p_{j}(\rho_{B};k)$ is the probability that Bob will detect Eve´s replacement photon in basis $j$ given that it was prepared in basis $k$, and we now limit ourselves to the cases where $\rho_A$, $\rho_B$ and $\rho_E$ are one of Alice, Bob or Eve's pre-defined measurement positions (A$x_1$,\dots, B$x_1$\dots, E$x_1$, \dots).  If Eve chooses the same measurement basis as both Alice and Bob, then 
she can go essentially undetected, since
\begin{equation}
P_{iii}(\rho_A,\rho_B,\rho_E)=R_{ii}(\rho_A,\rho_E)=P_{ii}(\rho_A,\rho_B),
\end{equation} 
which is the detection probability that Alice and Bob expect.  Here we have assumed Eve's best-case scenario, in which, given that she is completely aware of his detection system, she can replace Bob's photon in such a way that $p_{i}(\rho_B;i)=1$.  However, if Eve chooses the wrong basis, then, for the cases in which Alice and Bob expect a large detection probability $P_{ii}(\rho_A,\rho_B)$:
\begin{equation}
P_{iij}(\rho_A,\rho_B,\rho_E)=R_{ij}(\rho_A,\rho_E)p_{i}(\rho_B;i) \leq P_{ii}(\rho_A,\rho_B),
\end{equation} 
when $i \neq j$.  This follows from the fact that $R_{ij}(\rho_A,\rho_E) = P_{ij}(\rho_A,\rho_B) < P_{ii}(\rho_A,\rho_B)$ and $p_{i}(\rho_B;j) \leq 1$. In other words, 
the eavesdropping reduces the correlation between Alice and Bob's
measurements.  Similarly, if Eve intercepts the pair of
photons by another equally entangled pair, she is essentially playing the role of the source and the protocol is not affected \cite{bennett92}. If she replaces the pair of photons by
non- or less-entangled photons, then similarly the correlation between Alice and Bob's measurements is reduced and Eve's presence can be detected.
\par 
Assuming that Eve measures every one of Bob's photons in a basis ($x$ or $p$) chosen at random, the QBER for Eve's intercept-resend strategy is 
 \begin{equation}
 QBER =\frac{\sum\limits_{j=x,p}\sum\limits_{s,t=1,2}^{s\neq t}P({\mathrm{A}}j_s,{\mathrm{B}}j_t)+ \chi}
{\sum\limits_{i,j=x,p}\sum\limits_{s,t=1}^{2}P({\mathrm{A}}j_s,{\mathrm{B}}j_t)},
 \label{eq:qber3}
 \end{equation} 
 where 
  \begin{equation}
 \chi = \sum\limits_{j,k=x,p}^{j \neq k}\sum\limits_{s,t=1}^{2}p_{j}({\mathrm{B}}j_t;k)P({\mathrm{A}}j_s,{\mathrm{B}}j_t)
  \label{eq:qber4}
  \end{equation}
  is the error due to Eve's disturbance when she measures in the wrong basis.
  Here we have assumed that $R({\mathrm{A}}j_s,{\mathrm{E}}j_t) = P({\mathrm{A}}j_s,{\mathrm{B}}j_t)$. 
  \par

\par
From the argument above it is demonstrated that 
eavesdropping reduces the correlation between Alice and Bob's
measurements, and that this correlation is guaranteed by an EPR-like
state. The EPR character can be demonstrated, as in Ref. \cite{5},
by satisfying the inequality \cite{duan}:
\begin{equation}
\label{eq1} \Delta^2(Ax_i - Bx_i) \,\,\, \Delta^2(Ap_j + Bp_j) \leq
\frac{\hbar^2}{4}.
\end{equation}
To check for the presence of Eve, Alice and Bob can either check the correlations between $m$ pairs of their $N$ measurement results by calculating the QBER, or use these 
$m$ pairs to verify Eq. (\ref{eq1}). 
\section{Experiment}
\label{sec:experiment}

\begin{figure}[h]
\includegraphics*[width=8cm]{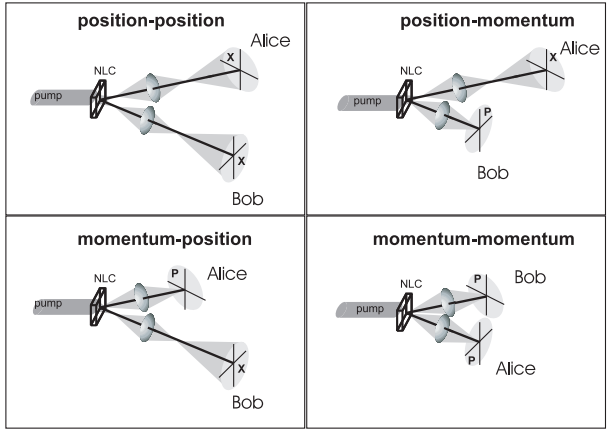}
\vspace{4mm}
\caption{Experimental arrangements used to test correlations in position and momentum degrees of freedom.  For position measurements a lens is used to image the face of the non-linear crystal (NLC) onto the detection plane, while for momentum measurements both the crystal face and the detection plane lie at the focal points of a lens (see text).} \label{fig2}
\end{figure}

We have performed measurements which demonstrate the security of a QKD protocol using position and momentum variables by testing the correlations
between position-position and  momentum-momentum coincidence detections, as well
the non-correlation between position-momentum and
momentum-position detections in a typical twin photon set-up.  The
experimental configuration is shown in Fig. \ref{fig2}. We use a
femto-second pulsed Ti-Sapphire laser doubled by a 2 mm long BBO
crystal, obtaining a violet beam with wavelength centered around
425 nm.  Using the violet beam to pump a 1 cm long Lithium
Iodate crystal, down-converted signal and idler photons were produced and
detected in different wavelengths with interference filters
centered in 890 nm with 10 nm bandwidth and 810 nm with 50 nm
bandwidth. We use avalanche photodiode single
photon counting modules equiped with short focal length lenses (often called objective lenses, because the focal lengths are
short, but they play the role of oculars) and a thin slit at the
entrance. The slits are oriented so that the horizontal dimension
is 3 mm and the vertical dimensions are 0.2 mm for position
measurements and  0.5 mm for momentum measurements.  When detection is performed in the position basis, the
crystal face is imaged by a 15 cm focal
length lens placed 20 cm from the crystal. The detectors are placed 60 cm from the lenses, giving a magnification factor of 3, which allows us  to image a narrow region with the 0.2 mm detection slits.  The pump beam is a Gaussian beam with spot size $\approx 2$ mm at the crystal face. For measurements in the momentum basis, a 15 cm focal
length lens is used and the detectors are placed 30 cm from the crystal face,
so that the crystal plane and the detection plane
coincide with the focal planes of the lens.  
\par

\begin{figure}[h]
\includegraphics*[width=8cm]{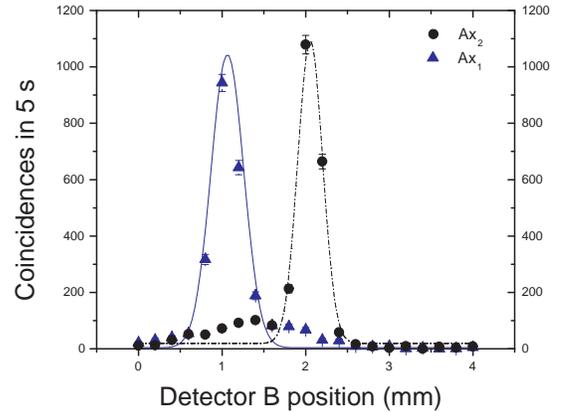}
\caption{Experimental results for $xx$ configurations.} \label{fig3a}
\end{figure}
\begin{figure}[h]
\includegraphics*[width=8cm]{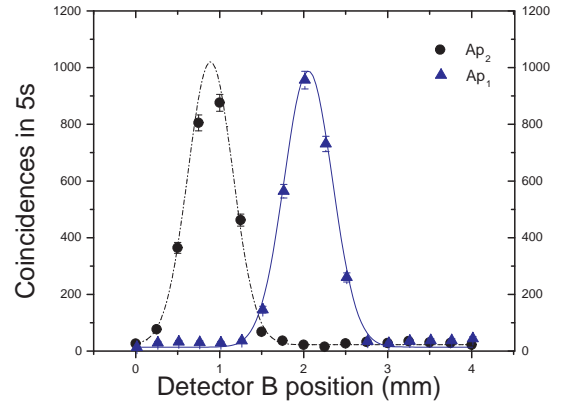}
\caption{Experimental results for $pp$ configurations.} \label{fig3b}
\end{figure}
\par
\begin{figure}[h]
\includegraphics[width=8cm]{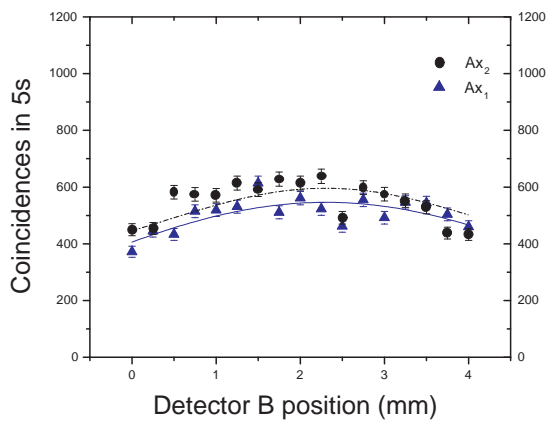}
\caption{Experimental results for $xp$ configurations.} \label{fig4a}
\end{figure}
\begin{figure}[h]
\includegraphics[width=8cm]{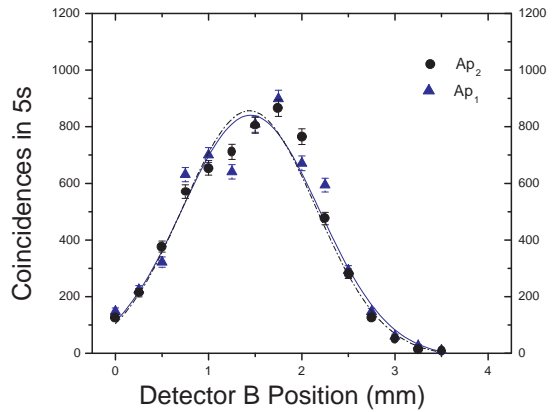}
\caption{Experimental results for $px$ configurations.} \label{fig4b}
\end{figure}

Results of the experimental investigations are displayed in Figs. \ref{fig3a} - \ref{fig4b}.  All curves are Gaussian fits and error bars represent errors due to Poissonian photon counting statistics.  In all measurements detector A was kept fixed while B was scanned linearly in the vertical direction.  Setting detectors A and B for position measurements, we expect to see a large
correlation in coincidence measurements. Fig. \ref{fig3a} shows the coincidence count rate when detector B is scanned
along the vertical axis and detector A
is fixed at position A$x_1$ (triangles) and also when A is fixed at
A$x_2$ (circles). The peaks of the coincidence
distributions are separated by about 1 mm, which is also the separation between A$x_1$
and A$x_2$, implying a good position correlation. The width of these
distributions is basically defined by the convolution of the
slit apertures in both detectors \cite{9}.  
\par
Fig. \ref{fig3b} shows the coincidence count rate when momentum
measurements are performed at both stations. Detector B was scanned while detector A was fixed at A$p_1$ (triangles) and  A$p_2$ (circles). A$p_1$ and A$p_2$ are separated by 1 mm and the
separation between the coincidence peaks is about 1 mm, showing a good momentum correlation. The widths of these
curves also depend on the overlap between the detection apertures, as the momentum measurement is actually mapped onto
measurements of the detector positions.  
\par
To show that the two-photon state is indeed EPR correlated, we calculate the variances in inequality (\ref{eq1}) using data from Figs. \ref{fig3a} and \ref{fig3b}, obtaining
\begin{eqnarray}
\label{eq2} \Delta^2(Ax_1 - Bx_1) & = (0.152 \pm 0.003)\mathrm{mm}^2, \nonumber \\
 \Delta^2(Ax_2 - Bx_2) & = (0.080 \pm 0.002)\mathrm{mm}^2, \nonumber \\
 \Delta^2(Ap_1 + Bp_1) & = (0.912 \pm 0.017) \hbar^2\, \mathrm{mm}^{-2}, \nonumber \\
\Delta^2(Ap_1 + Bp_1) & = (0.875 \pm 0.90) \hbar^2\, \mathrm{mm}^{-2}. 
\end{eqnarray}
These values give an average of $\Delta^2_x\,\Delta_p^2=(0.10 \pm 0.02) \hbar^2$,
which satisfies inequality (\ref{eq1}) by about $5$ standard
deviations.
\par
The security of our QKD protocol is based on a large correlation when measurements are performed in the same basis and 
low correlations when measurements are performed in complementary bases.
The measurements in Figs. \ref{fig3a} and \ref{fig3b} show that the correlation between
position-position measurements and momentum-momentum measurements
are very high. Now we are going to demonstrate that the
correlation between position-momentum and momentum-position is negligible. 
The coincidence profile for the $xp$ configuration is shown in Fig. \ref{fig4a}.  Here detector B is scanned in the $p$ configuration with detector A fixed at A$x_1$ (triangles)
and A$x_2$ (circles). 
There is no correlation between Alice and Bob's measurements in this case since the coincidence profile is approximately constant for all positions of detector B,  as expected from Eq. (\ref{eq:3}).  The slight ``enveloping" visible in the coincidence counts is due to a small modulation in the single counts as detector B is scanned (not shown).  Detector A positions A$x_1$ and A$x_2$ were chosen so that the coincidence rate is approximately the same for both cases.  We will show below that this is necessary to guarantee the security of the distributed key.   

Fig. \ref{fig4b} shows the coincidence profile for the $px$ configuration, where B is scanned in the $x$ basis with detector A fixed at A$p_1$ (triangles) and A$p_2$ (circles). As expected from Eq. (\ref{eq:4}), the coincidence count rate reproduces the transverse intensity profile of the pump field as a function of Bob's detector position only, which shows that in this case there is no correlation between Alice and Bob's measurements.   
\begin{table}
\caption{Coincidence counts for Alice and Bob's chosen detector positions.}
\begin{tabular}{ccccc}
Bob/Alice & A$x_1$ & A$x_2$ & A$p_1$ & A$p_2$ \\
\tableline
B$x_1$ & $943 \pm 31$ & $72 \pm 8$ & $700 \pm 26$& $655 \pm 26$\\
B$x_2$ & $67 \pm 8$ & $1079 \pm 33$ & $671 \pm 26$ & $765 \pm 26$ \\
B$p_1$ & $462 \pm 21$ & $492 \pm 22$ & $956 \pm 31$ & $22 \pm 5$ \\
B$p_2$ & $614 \pm 25$ & $591 \pm 24$ & $29 \pm 5$ & $876 \pm 30$ \\
\end{tabular}
\label{table1}
\end{table}

Using the results in Figs. \ref{fig3a} - \ref{fig4b},  Bob can choose his detector positions so that the security conditions discussed in section \ref{sec:theory} are best satisfied.  Examining the figures, it is most advantageous to define B$x_1=1$\,mm, 
and B$x_2=2$\,mm, B$p_1=1$\,mm and B$p_2=2$\,mm.
Table \ref{table1} shows the coincidence counts for every possible measurement at Alice and
Bob's stations, from which it can be seen that the correlation between measurements
in the same basis and the non-correlation between measurements in different bases.  There is a slight discrepency among the on-diagonal coincidence count rates.  A fine-tuning of the coincidence levels can always be acheived by using neutral filters in front of the detectors so that the on-diagonal terms are approximately equal.   
\par
From Table \ref{table1} and Eq. (\ref{eq:qber2}), the QBER in the absence of an eavesdropper would be $QBER_0=0.047 \pm 0.001$.  The QBER for $xx$ measurements is $QBER_0^{xx}=0.064 \pm 0.001$ 
and $QBER_0^{pp}=0.027 \pm 0.001$ for $pp$ measurements.  The security criterion for most QKD protocols is a QBER of less than about $0.15$ \cite{1}.  For a QBER above this limit it is impossible to establish a secret key, even with one-way error correction and privacy amplication. Since our QBER is much lower than this limit, it should be possible for Alice and Bob to establish a secure secret key.  
\par 
We can use the $xp$ and $px$ measurement results to predict the effect of an eavesdropper.  Using the results in Table \ref{table1} in Eq. (\ref{eq:qber3}), the estimated QBER if Eve were to measure every one of Bob's photons in a randomly chosen basis ($x$ or $p$) would be $QBER=0.296 \pm 0.001$ where we have assumed that $p_{j}({\mathrm{B}}j_t;k)=1/2$ for all $t$ and $j \neq k$ in Eq. (\ref{eq:qber4}).  In other words, we have assumed that if Eve measures in the wrong basis, she has a 50\% chance of sending the ``correct" replacement photon.  We see that Eve´s presence is clearly marked by an increase in the QBER.  
\section{Conclusion}
Recent work \cite{5} has shown that photon pairs created by spontaneous parametric down-conversion exhibit entanglement in position and momentum, of the sort originally proposed in the fundamental paper by Einstein-Podolsky-Rosen \cite{epr35}. Here we have extended these ideas by experimentally investigating the
implementation of a quantum key distribution protocol based on the
quantum correlations between position and momentum of entangled
photon pairs. In addition to interest due to the relation with historical debates on Quantum Theory, quantum key distribution based on position and momentum of photon pairs might offer some advantages. First, using these
degrees of freedom allows for the use of long non-linear
crystals, therefore opening the possibility of having really high
flux entangled-photon sources.  Second, it might be possible to extend these results to higher dimensional systems, which could be used in quantum communication protocols such as quantum bit commitment \cite{qbc}.
\begin{acknowledgements}
Financial support was provided by Brazilian agencies CNPq, PRONEX,
CAPES, FAPERJ, FUJB and the Milenium Institute for Quantum
Information.
\end{acknowledgements}
\vspace*{1.5cm}

* Corresponding author.\\ E-mail address: phsr@if.ufrj.br\\

\end{document}